\documentclass[pra,aps,amsmath,amssymb,twocolumn,showkeys,showpacs]{revtex4}

\newcommand{\ma}{\boldsymbol{a}}
\newcommand{\nn}{\boldsymbol{n}}
\newcommand{\qq}{\boldsymbol{q}}
\newcommand{\pp}{\boldsymbol{p}}
\newcommand{\mad}{\boldsymbol{a}^{\dagger}}
\newcommand{\bm}{\mathsf{B}}
\newcommand{\cm}{\mathsf{C}}
\newcommand{\ip}{\mathsf{\Pi}}
\newcommand{\nm}{\mathsf{N}}

\newcommand{\qp}{\mathsf{Q}}
\newcommand{\pq}{\mathsf{P}}
\newcommand{\am}{\mathsf{A}}
\newcommand{\ttm}{\mathsf{T}}
\newcommand{\amd}{\mathsf{A}^{\dagger}}
\newcommand{\niz}{\mathbf{0}}
\newcommand{\pen}{\openone}
\newcommand{\del}{\Delta_L}
\newcommand{\der}{\Delta_R}
\newcommand{\pxi}{\Psi_{{\xi}k}}
\newcommand{\uxi}{\Upsilon_{{\xi}k}}
\newcommand{\rt}{\mathrm{Tr}}
\newcommand{\hh}{\mathcal{H}}
\newcommand{\vac}{|n_0\rangle\langle{n}_0|}
\newcommand{\cp}{\mathsf{c}}

\newcommand{\rp}{\mathsf{r}}
\newcommand{\st}{\mathsf{s}}
\newcommand{\tf}{\tilde{f}}
\newcommand{\tpi}{\tilde{\varphi}}
\newcommand{\tif}{\tilde{f}}

\begin{document}
\clearpage
\preprint{}

\title{Entropic formulation of the uncertainty principle for the number and annihilation operators}
\author{Alexey E. Rastegin}
\affiliation{Department of Theoretical Physics, Irkutsk State University,
Gagarin Bv. 20, Irkutsk 664003, Russia}

\begin{abstract}
An entropic approach to formulating uncertainty relations for the
number-annihilation pair is considered. We construct some normal
operator that traces the annihilation operator as well as
commuting quadratures with a complete system of common
eigenfunctions. Expanding the measured wave function with respect
to them, one obtains a relevant probability distribution. Another
distribution is naturally generated by measuring the number
operator. Due to the Riesz-Thorin theorem, there exists a
nontrivial inequality between corresponding functionals of the above
distributions. We find the bound in this inequality and further
derive uncertainty relations in terms of both the R\'{e}nyi and
Tsallis entropies. Entropic uncertainty relations for continuous
distribution as well as relations for discretized one are
presented.
\end{abstract}

\maketitle

\pagenumbering{arabic}
\setcounter{page}{1}

\section{Introduction}

The concept of entropy has found use in many topics including a
quantification of uncertainty in quantum measurements. Entropic
uncertainty relations still attract much attention rather due to
its connection with some recent topics \cite{ww10}. The first
relation for the position--momentum pair in terms of the Shannon
entropies was derived by Hirschman \cite{hirs}. Hirschman's result
has later been improved \cite{beck}. A general statement of the
problem and utility of the entropic formulation have been
considered in the papers \cite{mamojka,deutsch}. An improvement of lower
bound from \cite{deutsch} had been conjectured by Kraus \cite{kraus} and later stated
by Maassen and Uffink \cite{maass}. The method of the paper
\cite{maass} is based on Riesz's theorem. Extensions of this
approach to mixed states and generalized measurements were considered in
\cite{krishna,rast102}. The entropic approach also allows to
formulate the uncertainty principle for trace-preserving quantum
operations \cite{rast104}.

There exist numerous approaches to the problem of finding a proper
expression for the number-phase uncertainty \cite{nieto,lynch}.
Both the well-defined Hermitian operator of phase and number-phase
uncertainty relation have been fit within the Pegg-Barnett
formalism \cite{barnett,radmore}. Various measures of quantum
phase uncertainty are compared in \cite{bfs93}. In the paper
\cite{lanz}, the uncertainty relation involving the number and
annihilation operators was proposed as an alternative to known
number-phase relations. Using the Pegg-Barnett formalism, the
number-phase uncertainty relations in terms of the Shannon entropy
were obtained \cite{abe92,gvb95,joshi}. For canonically conjugate
variables, the uncertainty relations were expressed in terms of
the Tsallis \cite{raja95} and R\'{e}nyi entropies \cite{zpv08}.
Mutually unbiased bases \cite{ivan92,sanchez93} and tomographic
processes \cite{mmanko09,mmanko10} have been examined within an
entropic approach as well.

In the present work, we aim to formulate the entropic uncertainty
principle for the number and annihilation operators. As entropic
measures, we will use both the R\'{e}nyi and Tsallis entropies.
Their definitions and some preliminary material are given in
Section \ref{ptlm}. The principal point of our approach is to
assign to the annihilation operator some normal operator in an
extended Hilbert space and related resolution of the identity.
Physically, this way implies the use of generalized measurement
with additional bosonic modes \cite{paris}. The simplest
construction of such a kind is described in Section \ref{genme}.
In Section \ref{rith}, we recall the Riesz-Thorin theorem and
derive an inequality between certain functionals of the
probability distributions generated by two measurements of
interest. In Section \ref{enun}, we obtain entropic uncertainty
relations for both the continuous distribution and distribution
with respect to finite bins. Section \ref{concl} concludes the
paper.

\section{Preliminaries}\label{ptlm}

In this section, the required facts are briefly outlined. Let
$\st=\{s_n\}$ be a probability distribution. For real $\alpha>0$
and $\alpha\not=1$, the R\'{e}nyi entropy of order $\alpha$ of
probability distribution $\st$ is defined by \cite{renyi}
\begin{equation}
R_{\alpha}(\st)=\frac{1}{1-\alpha}{\ }\ln\left(\sum\nolimits_n s_n^{\alpha}
\right) {\>}. \label{renent}
\end{equation}
For real $\alpha\not=1$, the Tsallis entropy of degree $\alpha$ is
defined by \cite{tsallis}
\begin{equation}
H_{\alpha}(\st)=\frac{1}{1-\alpha}{\,}\left(\sum\nolimits_n
s_n^{\alpha}-1\right)\equiv -\sum\nolimits_{n}
s_n^{\alpha}\ln_{\alpha}s_n \ , \label{tsaent}
\end{equation}
where $\ln_{\alpha}x\equiv\bigl(x^{1-\alpha}-1\bigr)/(1-\alpha)$
is the $\alpha$-logarithm. The Tsallis entropy of degree
$\alpha=2$ is related to the so-called degree of certainty which
is used for expressing complementarity relations for mutually
unbiased observables \cite{diaz}. A physical meaning of the
entropy (\ref{tsaent}) was the subject of some discussion (see
\cite{sattin} and references therein). In the context of
functional form (\ref{tsaent}), conditional and relative entropies
were introduced and analyzed \cite{sf06,rastjmp}. In the limit
$\alpha\to1$, both the entropies (\ref{renent}) and (\ref{tsaent})
coincide with the Shannon entropy, namely
\begin{equation}
R_1(\st)=H_1(\st)=-\sum\nolimits_n s_n\ln s_n
\ . \label{shent}
\end{equation}
Note that the entropies (\ref{renent}) and (\ref{tsaent}) are
both particular cases of more general class of unified entropies
\cite{hey06,rastjst}.

A quantum state is characterized by the probabilities of the
outcomes of every conceivable test \cite{peresq}. Consider
Hermitian operator $\bm=\sum_n b_n\ip_n$, where the $\ip_n$ is
orthogonal projector on the subspace corresponding to eigenvalue
$b_n$. The set $\{\ip_n\}$ of projectors obeys the completeness relation
$\sum_n \ip_n=\pen$, where $\pen$ is the identity. In the case of
continuous spectrum, the sums are replaced with the corresponding
integrals. If the state just prior to the measurement is described
by density operator $\rho$, then the outcome $b_n$ is obtained
with the probability $s_n=\rt\left(\rho\ip_n\right)$
\cite{peresq}. Its value is completely determined by the set
$\{\ip_n\}$ and density operator $\rho$. When the system was in
one of the eigenstates of $\bm$, the distribution is deterministic
and the entropies are zero. Therefore, for two and more
incompatible tests the sum of relevant entropies is nontrivially
bounded from below \cite{deutsch}. So, studies in the subject are
mainly devoted to obtaining such lower bounds for some
measurements of interest. As was shown by Maassen and Uffink
\cite{maass}, the Riesz theorem provides a unifying tool for these
aims.

There may be more than one ways to measure quantum uncertainty for
non-Hermitian operators. One of possible approach is to extend the
standard expression for variance. For arbitrary operator $\cm$, we
can define its left and right absolute values as
\begin{equation}
|\cm|_L:=\sqrt{\cm\cm^{\dagger}}
\ , \qquad
|\cm|_R:=\sqrt{\cm^{\dagger}\cm}
\ . \label{lefrav}
\end{equation}
We have $|\cm|_L=|\cm|_R$ if and only if $\cm$ is normal, i.e.
$\bigl[\cm,\cm^{\dagger}\bigr]=\niz$. Let us put the left and
right variances, $\del\cm\geq0$ and $\der\cm\geq0$, by
the equalities
\begin{align}
 & (\del\cm)^2=\langle{\,}|\cm|_L^2\rangle-\langle\cm\rangle\langle\cm^{\dagger}\rangle
\ , \label{lefrun1}\\
\qquad
 & (\der\cm)^2=\langle{\,}|\cm|_R^2\rangle-\langle\cm^{\dagger}\rangle\langle\cm\rangle
\ . \label{lefrun2}
\end{align}
For Hermitian operator $\bm$, these expressions are both reduced
to the regular expression
$(\Delta\bm)^2=\langle\bm^2\rangle-\langle\bm\rangle^2$. The
annihilation operator $\ma$ and the creation operator $\mad$ enjoy
$[\ma,\mad]=\pen$, whence $(\del\ma)^2=(\der\ma)^2+1$. In the
paper \cite{lanz}, the uncertainty relation is given in terms of
$\der\ma$ and $\Delta\nn$, where the number operator
$\nn=\mad\ma$. We could also use $\del\ma$ as a measure of
uncertainty of the operator $\ma$ in a given quantum state.

Another way is to relate the annihilation operator with a proper
resolution of the identity. Here we have to construct some
generalized measurement in an extended Hilbert space. Let us
introduce some notation. By $\{|n\rangle\}$ with integer $n\geq0$
we denote the number state basis in the Hilbert space $\hh$. Each
ket $|n\rangle$ obeys $\nn|n\rangle=n|n\rangle$. Putting the
dimensionless quadratures
\begin{equation}
\qq=\frac{\ma+\mad}{\sqrt{2}}   \ ,
\qquad \pp=\frac{\ma-\mad}{i{\,}\sqrt{2}}
\ , \label{defqd}
\end{equation}
we have $[\qq,\pp]=i{\,}\pen$ and $\ma=(\qq+i\pp)/\sqrt{2}$.
Considering two modes, we will denote their Hilbert spaces as
$\hh_x$ and $\hh_y$. The traces over $\hh_{x}$ and over
$\hh_{x}\otimes\hh_{y}$ will be written as $\rt_{x}(\centerdot)$
and $\rt_{xy}(\centerdot)$, respectively.

\section{Generalized measurement}\label{genme}

The aim is to build a normal operator $\am$ that act as $\ma$ on a
subset of states to be measured. For non-normal two-boson
operators, such a construction was analyzed in \cite{paris}.
The simplest way involves one additional mode. We denote the
measured bosonic mode by $x$ and the additional bosonic mode by
$y$. The $y$-mode is always prepared in the fixed initial state,
which is taken as some eigenstate $|n_0\rangle$ of the
number operator $\nn$. Consider two-boson operator
$\am=\ma_x\otimes\pen_y+\pen_x\otimes\mad_y$, acting in the
product space $\hh_x\otimes\hh_y$. It is important that this
operator is normal: $\bigl[\am,\amd\bigr]=\niz$. Writing
$\am=(\qp+i{\,}\pq)/\sqrt{2}$, the corresponding quadratures
\begin{align}
 & \qp=\frac{\am+\amd}{\sqrt{2}}=\qq_x\otimes\pen_y+\pen_x\otimes\qq_y
\ , \label{defqd1}\\
 & \pq=\frac{\am-\amd}{i{\,}\sqrt{2}}=\pp_x\otimes\pen_y-\pen_x\otimes\pp_y
\ , \label{defqd2}
\end{align}
are commuting operators, i.e. $[\qp,\pq]=\niz$.

On certain density operators, our extension leads to
the same average values for the quadratures and the same
measurement statistics for the number operator. For any density
operator $\rho$ on $\hh_x$, there hold
\begin{eqnarray}
&\rt_{xy}\bigl(\qp{\,}\rho\otimes\vac\bigr)=\rt_x(\qq_x\rho)
\ , \label{qpatra}\\
&\rt_{xy}\bigl(\pq{\,}\rho\otimes\vac\bigr)=\rt_x(\pp_x\rho)
\ , \label{qpatrb}\\
&\rt_{xy}\bigl(\am{\,}\rho\otimes\vac\bigr)=\rt_x(\ma_x\rho)
\ , \label{qpatrc}
\end{eqnarray}
We shortly write the above relations as
$\langle\qp\rangle=\langle\qq_x\rangle$,
$\langle\pq\rangle=\langle\pp_x\rangle$, and
$\langle\am\rangle=\langle\ma_x\rangle$. It is usually said that
the operator $\am$ ''traces'' the operator $\ma$ \cite{paris}.
We also define the operator $\nm=\nn_x\otimes\pen_y$ such that
\begin{align}
 & \rt_{xy}\bigl\{(|n\rangle\langle n|\otimes\pen_y)(\rho\otimes\vac)\bigr\}
=\langle n|\rho|n\rangle
\ , \label{sntat}\\
 & \rt_{xy}\bigl(\nm^{\nu}\rho\otimes\vac\bigr)=\rt_x(\nn_x^{\nu}\rho)
\ , \label{nstat}
\end{align}
for real power $\nu\geq0$. For other operators, the moments of
higher orders are not the same. Nevertheless, we have
\begin{align}
(\Delta\qp)^2&=(\Delta\qq_x)^2+n_{0}+\frac{1}{2}
{\>}, \label{uncon1}\\
\qquad
(\Delta\pq)^2&=(\Delta\pp_x)^2+n_{0}+\frac{1}{2}
{\>}, \label{uncon2}\\
(\Delta\am)^2&=(\del\ma_x)^2+n_{0}=(\der\ma_x)^2+n_{0}+1
{\,}. \label{uncon3}
\end{align}
It is important here that the above equalities do hold for any
initial state $\rho$ of the $x$-mode. Further, the two-boson
operator $\am$ can be measured practically. The measurement of
non-normal two-boson operators, which yield $\am$ as a particular
case, and its application in heterodyne detection are examined in
\cite{paris}.

We shall now find common eigenfunctions of the commuting operators
(\ref{defqd1}) and (\ref{defqd2}). In the coordinate
representation, the operator $\qp$ and $\pq$ are rewritten as
\begin{equation}
\qp=x+y=u \>, \quad
\pq=-{\,}i{\,}\frac{\partial}{\partial x}+i{\,}\frac{\partial}{\partial y}
=-2{\,}i{\,}\frac{\partial}{\partial v}
\>, \label{qpcoord}
\end{equation}
in terms of new variables $u=x+y$ and $v=x-y$, for which
$x=(u+v)/2$, $y=(u-v)/2$. A common eigenfunction $\uxi(u,v)$
obeys the equations
$$
u\uxi(u,v)=\xi\uxi(u,v) \ ,
\qquad
-2{\,}i{\,}\frac{\partial\uxi}{\partial v}=k\uxi(u,v)
\ ,
$$
where the eigenvalues $\xi$ of $\qp$ and $k$ of $\pq$ are
arbitrary real numbers. The solutions are written in a form
\begin{equation}
\uxi(u,v)=\frac{1}{2\sqrt{\pi}}{\>}\delta(u-\xi){\>}e^{ikv/2}
\label{pxisol}
\end{equation}
and satisfy the normalization condition
$$
\int du \int dv {\>} \uxi(u,v)^{*}{\>}\Upsilon_{{\xi'}k'}(u,v)=\delta(\xi-\xi'){\>}\delta(k-k')
\ .
$$
In terms of the variables $x$ and $y$, these eigenfunctions are
reexpressed as
\begin{align}
\pxi(x,y)&=\sqrt{2}{\>}\uxi(x+y,x-y)
\nonumber\\
&=\frac{1}{\sqrt{2\pi}}{\>}\delta(x+y-\xi){\>}e^{ik(x-y)/2}
\ , \label{reex}
\end{align}
since the Jacobian determinant $\partial(u,v)/\partial(x,y)=-2$.
So, the functions (\ref{reex}) also enjoy the desired property
$$
\int dx\int dy {\>}\pxi(x,y)^{*}{\>}\Psi_{{\xi'}k'}(x,y)=\delta(\xi-\xi'){\>}\delta(k-k')
\ .
$$
If the state of the $xy$-system right before the joint measurement
of $\qp$ and $\pq$ was $\Phi(x,y)$, then the probability density
is calculated by
\begin{equation}
w(\xi,k)=\left|\int dx \int dy {\ } \pxi(x,y)^{*}{\>}\Phi(x,y)\right|^2
\ . \label{wdens}
\end{equation}
The average values are calculated in the standard way, for
instance, $\langle\am\rangle=\bigl(\langle\qp\rangle+i\langle\pq\rangle\bigr)\big/\sqrt{2}$
and
\begin{equation}
\left\{
\begin{array}{c}
\langle\qp\rangle \\
\langle\pq\rangle
\end{array}
\right\}
=\int d\xi\int dk{\>\,}w(\xi,k)
\left\{
\begin{array}{cc}
\xi \\
k
\end{array}
\right\}
{\>}. \label{avpq}
\end{equation}
For $\alpha>0$ and continuous distributions, we will use a
norm-like $\alpha$-parametric functional defined as
\begin{equation}
\|w\|_{\alpha}= \left(\int d\xi \int dk {\ }w(\xi,k)^{\alpha}\right)^{1/\alpha}
{\,}. \label{walnor}
\end{equation}
This is actually a norm for $\alpha\geq1$. Denoting $n$th
eigenfunction of the number operator $\nn$ by $\varphi_n$, the
probability of obtaining  outcome in the $x$-mode is written as
$$
s_n=\int dx{\>}\varphi_n(x)^{*}\int dx'{\>}\varphi_n(x')
\int dy {\>}\Phi(x,y){\>}\Phi(x',y)^{*}
\ .
$$
We aim to relate the functionals $\|w\|_{\alpha}$ and
$\|\st\|_{\beta}=\Bigl(\sum_{n=0}^{\infty}
s_n^{\beta}\Bigr)^{1/\beta}$, whenever
$\Phi(x,y)=f(x){\,}\varphi_{n_0}(y)$.

\section{A consequence of the Riesz-Thorin theorem}\label{rith}

Recall some mathematical tools. By
$\cp=\bigl\{c_n\bigr\}_{n=0}^{\infty}$ we denote a sequence of
complex numbers, and by $l^{p}$ ($p\geq1$) the space of sequences
such that
\begin{equation}
\|\cp\|_{p}=\left(\sum\nolimits_{n-0}^{\infty}|c_n|^p\right)^{1/p}<\infty
\ . \label{normcp}
\end{equation}
For the complex-valued function $f(x)$ and $p\geq1$, the $p$-norm is defined as
\begin{equation}
\|f\|_{p}=\left(\int dx{\,}|f(x)|^p\right)^{1/p},
\label{normfp}
\end{equation}
with $\|f\|_{\infty}=\sup\{|f(x)|:{\,}x\in{\mathbb{R}}\}$ for the
special case $p=\infty$. By $L^p$ we denote the space of functions
such that $\|f\|_{p}<\infty$. We say that the linear mapping
$\ttm$ from $L^p$ to $L^q$ is {\it bounded} when its $(p,q)$-norm,
\begin{equation}
\|\ttm\|_{p,q}=\underset{f\neq0}{\sup}{\>}\frac{\|\ttm f\|_q}{\|f\|_p}
\ , \label{pqdf}
\end{equation}
is finite. In general, it is hard to find exactly values of
$(p,q)$-norms. Say, the answer is known for the Fourier transform
in both the discrete and continuous cases. In the former, it is
given by the Young-Hausdorff inequalities; in the latter, it has
been found by Beckner \cite{beck}. For arbitrary mapping, we can
employ the Riesz-Thorin interpolation theorem (see, e.g., theorem
1.1.1 in \cite{berg}). It should be noted, however, that this
theorem gives only an upper estimate on $(p,q)$-norms. Here the
conjugate indices $p,q\in[1;\infty]$ obey $1/p+1/q=1$.

{\bf The Riesz-Thorin theorem.} {\it Assume that $p_0\neq p_1$,
$\|\ttm\|_{p_0,q_0}=M_0$ and $\|\ttm\|_{p_1,q_1}=M_1$; then for
all $\theta\in(0;1)$,}
\begin{equation}
\|\ttm\|_{p_{\theta},q_{\theta}}\leq M_0^{1-\theta}M_1^{\theta}
\ , \label{risth}
\end{equation}
{\it where the indices $p_{\theta}$ and $q_{\theta}$ are put as}
\begin{equation}
\frac{1}{p_{\theta}}=\frac{1-\theta}{p_0}+\frac{\theta}{p_1}
\ , \qquad
\frac{1}{q_{\theta}}=\frac{1-\theta}{q_0}+\frac{\theta}{q_1}
\ . \label{conin}
\end{equation}

Suppose that $\bigl\{\varphi_n(x)\bigr\}_{n=0}^{\infty}$ is an orthonormal
set in the space $L^2$ and $n_0$ is a fixed number. To each wave
function $f(x){\,}\varphi_{n_0}(y)$ of the two modes with $f(x)\in L^2$,
we assign its decomposition with respect to the eigenfunctions
$\pxi(x,y)$, i.e.
\begin{equation}
f(x){\,}\varphi_{n_0}(y)=\int d\xi \int dk {\>\,}\tif(\xi,k){\,}\pxi(x,y)
\ . \label{decpx}
\end{equation}
Here the integral kernel $\tif(\xi,k)$ is represented as
\begin{align}
\langle\pxi,f{\,}\varphi_{n_0}\rangle&=
\frac{1}{\sqrt{2\pi}}\int dx \int dy{\ }\delta(x+y-\xi)
\nonumber\\
&\cdot{e}^{-ik(x-y)/2} f(x){\,}\varphi_{n_0}(y)
\ . \label{fordef}
\end{align}
It is clear that the transforms $\tpi_n(\xi,k)$ of the basis
functions $\varphi_n(x)$ form an orthonormal set, i.e.
$\langle\tpi_m,\tpi_n\rangle=\delta_{mn}$. Consider the linear map
$\ttm$ from $l^p$ to $L^q$ such that
\begin{equation}
\cp\mapsto\sum\nolimits_{n=0}^{\infty} c_n{\,}\tpi_n(\xi,k)
\label{tdf}
\end{equation}
for any sequence $\{c_n\}$. The right-hand side of (\ref{tdf}) can
be viewed as the transform $\tif(\xi,k)$ of the function
\begin{equation}
f(x)=\sum\nolimits_{n=0}^{\infty} c_n{\,}\varphi_n(x)
\ , \label{fcv}
\end{equation}
and we write $\ttm(\cp)=\tif$. It follows from the orthonormality
of the $\tpi_n(\xi,k)$'s that $\|\tif\|_2=\|\cp\|_2$, whence the
norm $\|\ttm\|_{2,2}=1$. Putting the quantity
\begin{equation}
\eta=\sup\bigl\{|\tpi_n(\xi,k)|:{\,}\xi,k\in{\mathbb{R}},{\,}n\in{\mathbb{N}}\bigr\}
\ , \label{etdef}
\end{equation}
we also have
$$
\|\tif\|_{\infty}=\underset{\xi,k\in{\mathbb{R}}}{\sup}\left|\sum\nolimits_{n=0}^{\infty}
c_n{\,}\tpi_n(\xi,k)\right|\leq\eta{\,}\sum\nolimits_{n=0}^{\infty} |c_n|
\ ,
$$
i.e. $\|\tif\|_{\infty}\leq\eta{\,}\|\cp\|_1$ and
$\|\ttm\|_{1,\infty}\leq\eta$. Using the Riesz-Thorin theorem with
the pairs $(p_0=2,q_0=2)$ and $(p_1=1,q_1=\infty)$, we obtain the
inequality
\begin{equation}
\|\ttm\|_{p,q}\leq\eta^{\theta}
\ . \label{risres}
\end{equation}
where the conjugate indices $p$ and $q$ are such that
$p=2/(1+\theta)<2$ and $q=2/(1-\theta)>2$. By $\theta=(2-p)/p$
and (\ref{pqdf}), the inequality (\ref{risres}) results in
\begin{equation}
\|\tf\|_q\leq\eta^{(2-p)/p}{\,}\|\cp\|_p
\ . \label{ris1}
\end{equation}
Assuming $1<p<2$ and $q>2$, we also have the relation
\begin{equation}
\|\cp\|_q\leq\eta^{(2-p)/p}{\,}\|\tf\|_p
\ , \label{ris2}
\end{equation}
which is derived from (\ref{ris1}) as follows. Let us write
$\gamma_n=|c_n|^q/c_n^{*}$ if $c_n\neq0$, and $\gamma_n=0$
otherwise. Then we have
\begin{equation}
\sum\nolimits_{n=0}^{\infty}|c_n|^q=\sum\nolimits_{n=0}^{\infty} \gamma_n^{*}{\,}c_n=
\langle\tilde{g},\tif\rangle \ , \label{tgft}
\end{equation}
where $\tilde{g}(\xi,k)=\sum_n\gamma_n{\,}\tpi_n(\xi,k)$ by definition. Due to the
H\"{o}lder inequality
$|\langle\tilde{g},\tif\rangle|\leq\|\tilde{g}\|_{q}{\,}\|\tif\|_{p}{\,}$,
there holds
\begin{equation}
\|\cp\|_{q}^{q}\leq \|\tilde{g}\|_{q}{\,}\|\tif\|_{p}
\leq\eta^{(2-p)/p}{\,}\|\boldsymbol{\gamma}\|_p{\,}\|\tif\|_{p}
\ , \label{cgam}
\end{equation}
where we have applied (\ref{ris1}) with $\tilde{g}$ and
$\boldsymbol{\gamma}$ instead of $\tif$ and $\cp$ respectively.
Transposing the factor
\begin{equation}
\|\boldsymbol{\gamma}\|_p=\left(\sum\nolimits_{n=0}^{\infty}|c_n|^{q}\right)^{1-1/q}
=\|\cp\|_{q}^{q-1}
\label{trans}
\end{equation}
from (\ref{cgam}), we finally obtain (\ref{ris2}).

For the number operator, the probabilities are
\begin{equation}
s_n=|c_n|^2=|\langle\varphi_n,f\rangle|^2
\ . \label{nopr}
\end{equation}
For the pre-measurement state
$\Phi(x,y)=f(x){\,}\varphi_{n_0}(y)$, the probability density
(\ref{wdens}) is represented as
\begin{equation}
w(\xi,k)=|\tif(\xi,k)|^2
\ . \label{rppd}
\end{equation}
We have $\|w\|_{\alpha}=\|\tif\|_{q}^2$ and
$\|\st\|_{\beta}=\|\cp\|_{p}^2$, where $\alpha=q/2$ and
$\beta=p/2$. Squaring (\ref{ris1}) and (\ref{ris2}), we obtain
\begin{align}
 & \|w\|_{\alpha}\leq\eta^{2(1-\beta)/\beta}\|\st\|_{\beta}
\ , \label{sir1}\\
 & \|\st\|_{\alpha}\leq\eta^{2(1-\beta)/\beta}\|w\|_{\beta}
\ , \label{sir2}
\end{align}
povided that $1/\alpha+1/\beta=2$ and $\alpha>1>\beta$. The
inequalities (\ref{sir1}) and (\ref{sir2}) directly lead to the
entropic uncertainty relation for the number and annihilation
operator.

\section{Entropic uncertainty relations}\label{enun}

Using the probabilities (\ref{nopr}), we take the sums in the
formulae (\ref{renent}) and (\ref{tsaent}) from $n=0$ up to
$n=\infty$. For the probability density (\ref{rppd}), we write
down
\begin{align}
 & R_{\alpha}(w)=\frac{1}{1-\alpha}{\ }\ln\|w\|_{\alpha}^{\alpha}
\ , \label{rtdef}\\
 & H_{\alpha}(w)=\frac{1}{1-\alpha}{\>}\bigl(\|w\|_{\alpha}^{\alpha}-1\bigr)
\ . \label{trdef}
\end{align}
The $n$th normalized eigenfunction of the operator $\nn$ is written as
\begin{equation}
\varphi_n(x)\equiv\langle x|n\rangle=\left(\sqrt{\pi}{\,}2^n n!\right)^{-1/2} e^{-x^2/2}{\,}h_n(x)
\ , \label{asef}
\end{equation}
where $h_n(x)$ is the $n$th Hermite polynomial \cite{radmore}. The
factor (\ref{etdef}) is the supremum of modulus of the quantity
\begin{align}
&\bigl\langle\pxi,\varphi_n\varphi_{n_0}\bigr\rangle
=\frac{1}{\sqrt{2\pi}}\int dx {\>}\varphi_n(x){\>}e^{-ikx/2}\int dy{\>}\varphi_{n_0}(y)
\nonumber\\
&\cdot\delta\bigl(y-(\xi-x)\bigr){\>}e^{iky/2}=\frac{1}{\sqrt{2\pi}}{\>\,}e^{ik\xi/2}\int dx{\ }\varphi_n(x)
\nonumber\\
&\cdot\varphi_{n_0}(\xi-x){\>}e^{-ikx}
=\frac{1}{\sqrt{2\pi}}{\>}\langle\varphi_n,\phi\rangle
\ . \label{scalpr}
\end{align}
Here we put the function
$\phi(x)=e^{ik\xi/2}{\,}\varphi_{n_0}(\xi-x){\>}e^{-ikx}$, for which
$\|\phi\|_2=\|\varphi_{n_0}\|_2=1$. Since the value
$\langle\varphi_n,\phi\rangle$ is $n$th coefficient in the
expansion
\begin{align}
 & \phi(x)=\sum\nolimits_{n=0}^{\infty} \langle\varphi_n,\phi\rangle{\>}\varphi_n(x)
\ , \label{hexpan1}\\
 & \sum\nolimits_{n=0}^{\infty} |\langle\varphi_n,\phi\rangle|^2=\|\phi\|_2^2=1
\ , \label{hexpan2}
\end{align}
the modulus of each coefficient does not exceed one. Combining
this with (\ref{scalpr}) gives
\begin{equation}
{\sup}\left\{\bigl|\bigl\langle\pxi,\varphi_n\varphi_{n_0}\bigr\rangle\bigr|:{\,}
\xi,k\in{\mathbb{R}},{\,}n\in{\mathbb{N}}\right\}\leq\frac{1}{\sqrt{2\pi}}
\ . \label{etub}
\end{equation}
The case $n_0=0$, when no photons are initially being in the
$y$-mode, is easier to realize experimentally. For $n=0$,
direct calculations give
\begin{align}
&\frac{1}{\sqrt{2\pi}}{\>\,}e^{ik\xi/2}\int dx{\>}\varphi_0(x){\>}\varphi_0(\xi-x){\>}e^{-ikx}=
\label{in00}\\
&\frac{1}{\sqrt{2\pi}}{\>\,} \frac{1}{\sqrt{\pi}}{\>\,}e^{-\xi^2/4}
\int dz {\>}e^{-z^2-ikz}=\frac{1}{\sqrt{2\pi}}{\>\,}e^{-(\xi^2+k^2)/4}
\ , \nonumber
\end{align}
where $z=x-\xi/2$. The supremum of (\ref{in00}) with respect to
$k$ and $\xi$ is equal to $(2\pi)^{-1/2}$, that is exactly
$\eta=(2\pi)^{-1/2}$.

By (\ref{etub}), we rewrite the inequalitie (\ref{sir1}) and (\ref{sir2}) as
\begin{align}
 & \|w\|_{\alpha}\leq(2\pi)^{-(1-\beta)/\beta}\|\st\|_{\beta}
\ , \label{wrrw1}\\
 & \|\st\|_{\alpha}\leq(2\pi)^{-(1-\beta)/\beta}\|w\|_{\beta}
\ , \label{wrrw2}
\end{align}
where $1/\alpha+1/\beta=2$ and $\alpha>1>\beta$. The
uncertainty relation in terms of the R\'{e}nyi entropies is given by
\begin{equation}
R_{\alpha}(w)+R_{\beta}(\st) \geq \ln2\pi
\ , \label{eqreip}
\end{equation}
under the condition $1/\alpha+1/\beta=2$. Taking the logarithm of (\ref{wrrw1})
and doing some algebra, we just obtain
\begin{equation}
\frac{1-\alpha}{\alpha}{\ }\frac{\beta}{1-\beta}{\ }R_{\alpha}(w)\leq-\ln2\pi+R_{\beta}(\st)
\ . \label{aabb}
\end{equation}
Here the multiplier of $R_{\alpha}(w)$ is equal to $(-1)$. Hence
we get (\ref{eqreip}) in the case when the entropy of the
distribution $w(\xi,k)$ has larger order. Otherwise, we repeat the
same with the inequality (\ref{wrrw2}). The derivation is slightly
more complicated with the Tsallis entropies. We claim that
\begin{equation}
H_{\alpha}(w)+H_{\beta}(\st)\geq\ln_{\mu}2\pi
\ , \label{eqtsip}
\end{equation}
where $1/\alpha+1/\beta=2$ and $\mu=\max\{\alpha,\beta\}$. Indeed,
we rewrite the left-hand side of (\ref{eqtsip}) as the function
\begin{equation}
g(t,\tau)=\frac{t-1}{1-\alpha}+\frac{\tau-1}{1-\beta}
\label{sumhh}
\end{equation}
in terms of the variables $t=\|w\|_{\alpha}^{\alpha}=\int d\xi
\int dk {\>}w(\xi,k)^{\alpha}$ and
$\tau=\|\st\|_{\beta}^{\beta}=\sum_{n}s_n^{\beta}$. Assuming
$\alpha>1>\beta$, there holds $t\leq1$ and $\tau\geq1$ due to
$w(\xi,k)\leq\eta^2<1$ and the normalization relation. Adding
(\ref{wrrw1}) in the form
$(2\pi)^{1-\beta}t^{\beta/\alpha}\leq\tau$, we want to minimize
$g(t,\tau)$ under the above conditions. In view of Appendix of the
paper \cite{rast104}, the minimum is reached for
$t_0=(2\pi)^{1-\alpha}$, $\tau=1$ and equal to
$g(t_0,1)=\ln_{\alpha}2\pi$. In the same manner, we  use
(\ref{wrrw2}) and resolve the case when the Tsallis entropy of the
distribution $\st$ has larger degree. It is proper that the entropic
uncertainty relations (\ref{eqreip}) and (\ref{eqtsip}) coincide in the limit
$\mu\to1$.

The relations (\ref{eqreip}) and (\ref{eqtsip}) involve the
continuous distribution $w(\xi,k)$. We can also obtain similar
relations in terms of probabilities
\begin{equation}
r_{lm}=\int_{\xi_l}^{\xi_{l+1}} d\xi \int_{k_m}^{k_{m+1}} dk{\>\,} w(\xi,k)
\ , \label{sbin}
\end{equation}
where $\{\xi_l\}$ is a partition of the $\xi$-axis and $\{k_m\}$
is a partition of the $k$-axis. Let $\Delta\xi=\max\Delta\xi_l$,
$\Delta k=\max\Delta k_{m}$ be maximal sizes of bins on these
axes. By theorem 192 of the book \cite{hardy} for integral means,
we have
\begin{align}
& \left(\frac{1}{\Delta\xi_l\Delta k_{m}}{\>} \int_{\xi_l}^{\xi_{l+1}}
d\xi \int_{k_m}^{k_{m+1}} dk{\>\,} w(\xi,k)\right)^{\!\alpha}
\left\{
\begin{array}{cc}
\leq, & \alpha>1 \\
\geq, & \alpha<1
\end{array}
\right\}
\nonumber\\
&{\ }\frac{1}{\Delta\xi_l\Delta k_m}{\>} \int_{\xi_l}^{\xi_{l+1}} d\xi \int_{k_m}^{k_{m+1}} dk
{\>\,} w(\xi,k)^{\alpha}
\ , \label{aconv}
\end{align}
whence $\bigl(\Delta\xi\Delta
k\bigr)^{(1-\alpha)/\alpha}\|\rp\|_{\alpha}\leq\|w\|_{\alpha}$ for
$\alpha>1$, and $\|w\|_{\beta}\leq\bigl(\Delta\xi\Delta
k\bigr)^{(1-\beta)/\beta}\|\rp\|_{\beta}$ for $\beta<1$. The last
inequalities are obtained by summing the inequality (\ref{aconv})
and raising these sums to the powers $1/\alpha$ and $1/\beta$
respectively. Since $(\alpha-1)/\alpha=(1-\beta)/\beta$, we then
deduce from (\ref{wrrw1}) and (\ref{wrrw2}) that
\begin{align}
&\|\rp\|_{\alpha}\leq\left(\frac{\Delta\xi\Delta k}{2\pi}\right)^{(1-\beta)/\beta}\|\st\|_{\beta}
\ . \label{dedc1}\\
&\|\st\|_{\alpha}\leq\left(\frac{\Delta\xi\Delta k}{2\pi}\right)^{(1-\beta)/\beta}\|\rp\|_{\beta}
\ . \label{dedc2}
\end{align}
Only for $\Delta\xi\Delta k<2\pi$ these inequalities contain a
nontrivial constraint ($\|\centerdot\|_{\alpha}\leq1$ by
$\alpha>1$, $\|\centerdot\|_{\beta}\geq1$ by $\beta<1$). The
uncertainty relations are now written as
\begin{align}
& R_{\alpha}(\rp)+R_{\beta}(\st) \geq \ln\!\left(\frac{2\pi}{\Delta\xi\Delta k}\right)
\ , \label{eqreip1} \\
& H_{\alpha}(\rp)+H_{\beta}(\st) \geq \ln_{\mu}\!\left(\frac{2\pi}{\Delta\xi\Delta k}\right)
\ , \label{eqtsip1}
\end{align}
where $1/\alpha+1/\beta=2$ and $\mu=\max\{\alpha,\beta\}$.

The above entropic uncertainty relations can all be extended to
mixed initial states of the $x$-mode. For the density operator
$\rho=\sum_{\lambda}\lambda{\,}|\lambda\rangle\langle\lambda|$, we
put the functions $f_{\lambda}(x)=\langle x|\lambda\rangle$ and
the total probability distributions
\begin{equation}
w(\xi,k)=\sum\nolimits_{\lambda}\lambda{\,}w_{\lambda}(\xi,k)
\ , \qquad
s_n=\sum\nolimits_{\lambda}\lambda{\>}s_n^{(\lambda)}\ .
\label{wlm}
\end{equation}
Here $s_n^{(\lambda)}=|\langle\varphi_n,f_{\lambda}\rangle|^2$,
$w_{\lambda}(\xi,k)=|\tif_{\lambda}(\xi,k)|^2$, and the
$\tif_{\lambda}(\xi,k)$ is defined by (\ref{fordef}) with
$f_{\lambda}(x)$ instead of $f(x)$. For each $\lambda$, there hold
the inequalities (\ref{wrrw1}) and (\ref{wrrw2}) with
$w_{\lambda}(\xi,k)$ and $s_n^{(\lambda)}$. By the Minkowski
inequality, the same inequalities are valid for the total
distributions (\ref{wlm}). We refrain from presenting the details
here (cf. equation (3.9) of \cite{rast102}). So the uncertainty
relations (\ref{eqreip}), (\ref{eqtsip}), (\ref{eqreip1}), and
(\ref{eqtsip1}) can all be derived.

\section{Conclusions}\label{concl}

We have considered an approach to formulate number-annihilation
uncertainty by means of entropic measures. To the annihilation
operator we have assigned the normal two-boson operator enjoying
the same average values for measured density matrices. Its
measurement can be implemented in practice. The corresponding
Hermitian quadratures are commuting with a common set of
eigenfunctions. The decomposition of any wave function of interest
with respect to the common eigenfunctions allows to obtain a
proper probability distribution. The non-trivial entropic bounds
for two generated probability distributions have been derived in
terms of both the R\'{e}nyi and Tsallis entropies. Entropic
uncertainty relations of a state-independent form are written down
for the continuous distribution (see (\ref{eqreip}) and
(\ref{eqtsip})) as well as for discretized one (see
(\ref{eqreip1}) and (\ref{eqtsip1})). The discretized distribution
is taken with respect to some partitions on the corresponding
axes. Nontrivial entropic bounds are given for the case, when any
product of two bins is less than $2\pi$. The latter concurs with
the fact that one quantum degree of freedom occupies a
(dimensionless) phase cell by size at least $2\pi$. The same
entropic bounds remain valid for an impure measured state.

\acknowledgments

I am grateful to S. Abe for bringing \cite{abe92} to my attention and to anonimous referee for constructive criticism. This work was supported in a part by the Ministry of Education and Science of the Russian Federation under grants no. 2.2.1.1/1483, 2.1.1/1539.

\end{document}